\documentclass[twocolumn,showpacs,prb,preprintnumbers,amsmath,amssymb]{revtex4}
\usepackage[latin1]{inputenc}
\usepackage{graphicx}% Include figure files
\usepackage{dcolumn}% Align table columns on decimal point
\usepackage{bm}% bold math

\newcommand{\si}[1]{\text{sgn}(}

\begin{document}

\title{Breakdown of Fermi liquid behavior near the hot spots in a two-dimensional model: A 
two-loop renormalization group analysis}
\author{Vanuildo S. de Carvalho$^{1}$}
\author{Hermann Freire$^{1,2}$}
\email{hfreire@mit.edu}
\affiliation{$^{1}$Instituto de Física, Universidade Federal de Goiás, 74.001-970, Goiânia-GO, Brazil}
\affiliation{$^{2}$Department of Physics, Massachusetts Institute of Technology, Cambridge, Massachusetts 02139, USA}

\date{\today}

\begin{abstract}
Motivated by a recent experimental observation of a nodal liquid on both single crystals and thin films of Bi$_2$Sr$_2$CaCu$_2$O$_{8+\delta}$ 
by Chatterjee \emph{et al.} [Nature Physics \textbf{6}, 99 (2010)], we perform a field-theoretical renormalization 
group (RG) analysis of a two-dimensional model consisting of 
eight points located near the ``hot spots'' on the Fermi surface
which are directly connected by spin density wave ordering wave vector. We derive RG equations up to two-loop order describing the flow of
renormalized couplings, quasiparticle weight, several order-parameter response functions, and 
uniform spin and charge susceptibilities of the model.
We find that while the order-parameter susceptibilities investigated here become non-divergent at two loops, 
the quasiparticle weight vanishes in the low-energy limit, indicating a breakdown of Fermi liquid behavior at this RG level.
Moreover, both uniform spin and charge
susceptibilities become suppressed in the scaling limit which indicate gap openings in both spin and charge excitation spectra of the model. 
\end{abstract}

\pacs{74.20.Mn, 74.20.-z, 71.10.Hf}

\maketitle

\section{Introduction}

The nature of the pseudogap state which manifests itself in underdoped cuprates continues to generate both
interest and controversy in the field of high-T$_{c}$ superconductivity.
This is because the proper understanding of this phase turns out to be crucial for comprehending the underlying
mechanism of superconductivity displayed by these materials \cite{Lee,Anderson}.
In recent years, angle-resolved photoemission spectroscopy (ARPES) shed new light on this problem.
Experiments \cite{Campuzano} performed at very low temperatures by Chatterjee \emph{et al.} on both single crystals and thin films of Bi$_2$Sr$_2$CaCu$_2$O$_{8+\delta}$ gave important 
evidence pointing to the existence of an electronic state
-- located in between the antiferromagnetic (AF) and $d$-wave superconducting (SC) phases -- whose excitation spectrum becomes zero only at the so-called nodal points 
(i.e. in the direction along the line connecting $\Gamma=(0,0)$ and $M=(\pi,\pi)$-points in momentum space). This state has thus been called a nodal liquid \cite{Anderson2,Nayak,Tesanovic}. This electronic 
state is insulating but with a $d$-wave gap structure
and exhibits non-Fermi liquid behavior around the so-called ``hot spots'' (i.e. points in momentum space where the AF zone boundary intersects the underlying Fermi surface of the system).

A minimal electronic model which potentially contains
the rich phenomenology of the phase diagram displayed by the cuprates is the two-dimensional
(2D) Hubbard model on a square lattice. This conjecture, originally put forward by Anderson \cite{Anderson} in the mid 80's, sparked among other things a profound reexamination 
of the foundations of Landau Fermi-liquid theory and, particularly, the precise conditions in which this standard theory breaks down in some 2D strongly correlated models. 
Indeed, soon after the experimental discovery of the cuprate superconductors, Anderson proposed that the ground state of the 2D Hubbard model slightly away from half-filling 
is given by an insulating spin liquid (ISL) which has
no broken symmetries down to zero temperature due to strong quantum fluctuations and also exhibits no well-defined quasiparticle excitations at low energies\cite{Lee2,Balents}. 
This question, however, remains unresolved to this date. Despite that, these ISL states are very interesting from
a theoretical viewpoint due to the fact that, once they are lightly doped with holes, they indeed give
rise to a singlet $d$-wave superconducting phase at mean-field level\cite{Kotliar} in agreement with experimental observation.

On the other hand, for a long time, there has not been a rigorous proof of the stability of this ISL state in realistic
microscopic models. Part of this imbroglio owes to the fact that, at the present time, there is no general analytical or numerical technique
that allows one to solve in a exact (or nearly exact) way such strongly correlated models in 2D.
In the last years, important advances have been made with improvements in both analytical methods (see, e.g., Refs. \cite{Senthil,Sondhi,Kitaev,Senthil2,Wen,Wen2}) and
also numerical techniques\cite{Muramatsu,White,Balents2} which, 
supplemented by groundbreaking experimental results\cite{YoungLee,Kanoda,YoungLee2}, provided strong evidence for the existence of these new quantum states in
some realistic physical situations.

From a weak-to-moderate coupling perspective, renormalization group (RG) methods remain one of the most powerful tools to
attack these problems in view of its unbiased nature \cite{Shankar,Kopietz,Metzner_Review}. They
have been applied to many electronic models ranging from simpler 2D models\cite{Dzyaloshinskii,Lederer,Gonzalez,Rice,Rice2,Binz,Ferraz} which focus on the dominant role of
scattering processes involving Fermi surface regions of the model to the fully 2D
Hubbard model defined on different lattice types. 
Regarding the 2D Hubbard model on a square lattice for weak-to-moderate couplings, for instance, some of these
RG works successfully reproduced both at
one-loop\cite{Zanchi,Halboth,Honerkamp} and two-loop orders\cite{Freire,Katanin} an AF phase in the
model near half-filling, the onset of a $d$-wave singlet SC phase
away from half-filling, and an additional electronic phase reminiscent of the pseudogap state\cite{Kataninetal,Rohe,Freire} which interpolates between these two phases. This agrees qualitatively with
the physics displayed by the cuprate superconductors and gives further support to the
point of view that this 2D model might indeed capture several important features of
these strongly-correlated materials.

Motivated also by those theoretical findings, we present in this paper a two-loop field-theoretical RG calculation of the 
renormalized couplings, the quasiparticle weight, several order-parameter response functions, and the uniform spin and charge susceptibilities
of a 2D fermionic model consisting of eight points located near the ``hot spots'' on the Fermi surface
which are directly connected by spin density wave (SDW) ordering wave vector.
A variation of this model was recently investigated in the literature using RG methods by Abanov and Chubukov\cite{Abanov,Abanov2}, Metlitski and Sachdev\cite{Sachdev},
and Efetov \emph{et al.}\cite{Efetov},
in which the corresponding high-energy fermions are partially integrated out in the system such that they arrive at a low-energy effective theory
involving the bosonic SDW order parameter coupled to the fermionic excitations near the ``hot spots''. They found interesting
renormalizations of several parameters of the model and also a clear breakdown of Fermi liquid behavior near
these Fermi points. By contrast, our approach in the present paper follows closely the spirit of the conventional RG strategy\cite{Shankar}, in which
high-energy degrees of freedom are successively integrated out 
to derive an effective description in terms of renormalized low-energy fermions. This approach can be found, for instance,
in a pioneering analysis performed by Furukawa and Rice\cite{Rice} who first discussed a very similar model within a purely fermionic RG scheme  
up to one-loop order. As a consequence, these authors found, for weak Hubbard-like initial conditions, a 
RG flow of the renormalized interactions towards strong coupling at low energies, 
which seems to suggest some limitations of the one-loop RG scheme in order to have complete access to the infrared regime of this system.
In our view, this fact also points to the importance of higher-order quantum corrections in
the full description of the low-energy dynamics of this
2D model. In this respect, our present work takes seriously this observation and, for this reason, it represents a step
forward in this direction.

This paper is organized as follows. In Sect. II, the 2D fermionic model consisting of eight points located near the ``hot spots'' on the Fermi surface
that we wish to study is introduced. Next (in Sect. III), we explain the 
field-theoretical RG methodology and show how to implement this method up to two loops in order to describe this model. 
In this part, we will choose to present this methodology in a concise way
since the field-theoretical RG scheme up to two-loop order was
already described in detail by one of us in the context of
another 2D fermionic model elsewhere \cite{Freire2}. The RG flow
equations are derived analytically and solved numerically in Sect. IV and then we proceed to discuss our main results.
Lastly, Sect. V is devoted to our conclusions.

\section{Model}

We begin our analysis with a general two-dimensional energy dispersion given, for instance,
by $\xi_{\mathbf{k}}=-2t(\cos(k_x)+\cos(k_y))-4t'\cos(k_x)\cos(k_y)-\mu$
with $t$ being the nearest neighbor hopping, $t'$ the next-nearest neighbor hopping amplitude and $\mu$ the
chemical potential. For the cuprates, the appropriate choice of parameters is $t'=-0.3t$, which results, at low hole doping, in the curved Fermi
surface (FS) shown in Fig. 1. This FS intersects the antiferromagnetic Brillouin zone at eight points (i.e., the ``hot spots'').
These points are connected through Umklapp processes which leads us to examine the RG equations for all the
coupling constants near these points.

\begin{figure}[t]
 \includegraphics[height=2.5in]{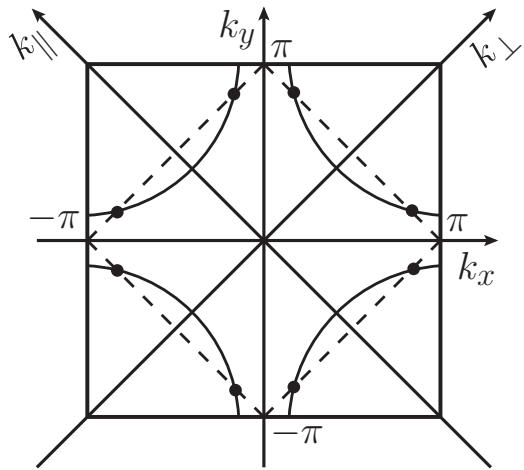}
 \caption{The FS of the 2D fermionic model consisting of eight points located near the ``hot spots'' on the Fermi surface
 which are directly connected by spin density wave (SDW) ordering wave vector.}
\end{figure}

The eight points around the ``hot spots'' are displayed in Fig. 1. If we rotate the momentum axes $(k_x,k_y)$ by $45^{\circ}$ degrees,
we can define the new axes $(k_{\parallel},k_{\perp})$, where the momenta $k_{\parallel}$ and $k_{\perp}$ refer to momentum parallel
and normal to the FS, respectively. Besides, since we
will be interested only in the universal quantities of this
model, we shall linearize the energy dispersion around the FS as $\xi_{\mathbf{k}}\approx v_{F,\perp} (k_{\parallel}) (|k_{\perp}|-k_{F,\perp})$
with the component of the Fermi velocity given by $v_{F,\perp} (k_{\parallel})=|\nabla_{(k_{\parallel},k_{\perp})}\xi_{\mathbf{k}}|_{k_{\perp}=k_{F,\perp}}|
\sin{\alpha}$, where 
$k_{F,\perp}$ is the perpendicular component of the Fermi momentum at the hot spots and $\alpha$ is the angle of the Fermi velocity on 
the rotated momentum axes. In terms of the parameters of 
the two-dimensional energy dispersion, the perpendicular component of the Fermi velocity is given by

\vspace{-0.2cm}

\begin{eqnarray}
v_{F,\perp} (k_{\parallel})&=&\bigg\{8\sin^{2}\left(\frac{k_{\parallel}}{\sqrt{2}}\right)\left[t\cos\left(\frac{k_{F,\perp}}{\sqrt{2}}\right)
-t'\cos\left(\frac{k_{\parallel}}{\sqrt{2}}\right)\right]^{2}\nonumber\\
&+&(k_{\parallel}\leftrightarrow k_{F,\perp})\bigg\}^{1/2}\sin{\alpha},
\end{eqnarray}

\noindent and $k_{F,\perp}$ is given by

\vspace{-0.4cm}

\begin{eqnarray}
k_{F,\perp}=\pm\sqrt{2}\arccos\left\{\frac{4t\cos\left(\frac{k_{\parallel}}{\sqrt{2}}\right)\pm\sqrt{\Delta}}{8t'\left[1+\cos^{2}\left(\frac{k_{\parallel}}{\sqrt{2}}\right)\right]}\right\}.
\end{eqnarray}

\noindent where $\Delta=16t^{2}\cos^{2}(k_{\parallel}/\sqrt{2})+16t'[1+\cos^{2}(k_{\parallel}/\sqrt{2})][\mu+4t'\sin^{2}(k_{\parallel}/\sqrt{2})]$.
Both the momenta parallel to the FS and perpendicular to the FS are restricted to the interval $[-k_{c},k_{c}]$, where $k_c$
essentially determines the ultraviolet (UV) momentum cutoff in our theory. This implies also an energy cutoff which is given by $\Lambda_0=2v_{F}k_{c}$ which
we choose to be equal to the full bandwidth of problem, i.e. $\Lambda_0=8t$.

If we use a coherent-state functional integral
representation of the resulting Hamiltonian, the model at $T = 0$ and constant
chemical potential $\mu = E_F$ becomes described by the partition function $\mathcal{Z}=\int \mathcal{D}[\overline{\psi},\psi] \exp({i\int_{-\infty}^{\infty} dt\, L[\overline{\psi},\psi]})$ with the
Lagrangian $L = L_{0} + L_{int}$ given by

\begin{align}\label{lagrangian}
&L=\sum_{\mathbf{k},\sigma}\overline{\psi}_{\sigma}(\mathbf{k})\left(i\partial_t-\xi_{\mathbf{k}}\right)\psi_{\sigma}(\mathbf{k})\nonumber\\
&-\sum_{i}\sum_{\substack{{\mathbf{k_1,k_2,k_3}} \\ {\sigma,\sigma'}}}
g_{i,B}\,\overline{\psi}_{\sigma}(\mathbf{k_4})\overline{\psi}_{\sigma'}(\mathbf{k_3})\psi_{\sigma'}(\mathbf{k_2})\psi_{\sigma}(\mathbf{k_1})
\end{align}

\noindent where $\mathbf{k_4}=\mathbf{k_1}+\mathbf{k_2}-\mathbf{k_3}$ and the volume $V$ has been set equal to unity. The Grassmann fields
$\overline{\psi}_{\sigma}(\mathbf{k})$ and $\psi_{\sigma}(\mathbf{k})$ are associated, respectively, to the creation and annihilation operators
of excitations lying in the vicinity of the hot spots with momentum $\mathbf{k}$ and spin projection $\sigma$. The index $i$ runs over all possible interaction processes of
the model that produce logarithmic divergences within perturbation theory, i.e., $i=1,2,3,1c,2c,1x,2x,1s,1r,3x,3p,3t$ (for details
of all the couplings taken into account, see Fig. 2). In this way, in order to keep a close connection with other RG works in the literature,
we are following a ``g-ology'' notation, adapted of course to our 2D problem at hand.
The Lagrangian of Eq. (\ref{lagrangian}) therefore defines our bare
quantum field theory model which is regularized in the UV by the cutoff $k_c$ mentioned above.

\section{Two-loop RG Methodology}

The methodology of our RG scheme follows closely the standard field-theoretical approach\cite{Peskin}, which was
also explained in full detail in a previous paper published by one of us\cite{Freire2}. If one applies a naive perturbation theory for the present model, 
divergences (or non-analyticities) emerge in the 
low-energy limit at the calculation of several
important quantities of the model such as one-particle irreducible four-point vertices, self-energy, and linear response functions. This result 
normally implies that the bare perturbation theory setup is not appropriate for this case, since it is known to be written
in terms of the microscopic parameters and not the low-energy quantities of the model. We circumvent this problem
by rewriting all the bare parameters of the theory in terms of the corresponding renormalized ones plus additional counterterms. 
The main role of these counterterms is to regularize 
the theory at a floating RG scale $\Lambda$ and, for this reason, they must be calculated order by order in perturbation theory. 
By doing this, the newly-constructed renormalized perturbation theory becomes a well-defined expansion in terms of the 
renormalized couplings and, in this way, its predictions can be compared to experiments. Since this program is successfully accomplished here, the 
field theory model is said to be renormalizable.

\begin{figure}[t]
 \includegraphics[height=2.9in]{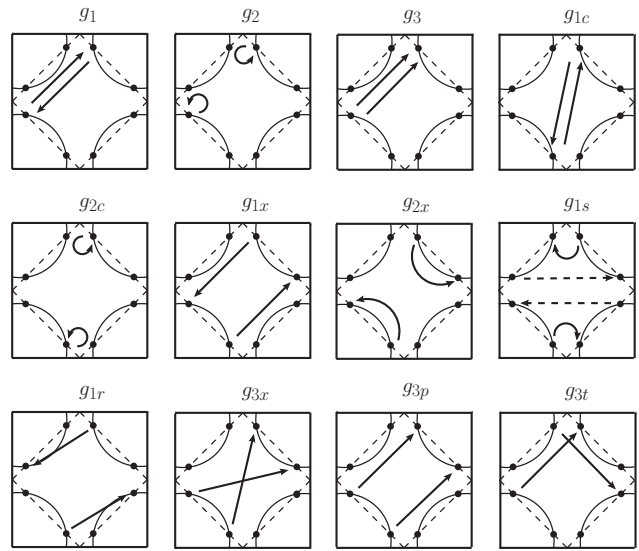}
 \caption{The relevant couplings of the field theory model. Here we use a ``g-ology'' notation adapted to our 2D problem at hand.}
 \label{fig:Interactions}
\end{figure}

At two-loop RG level, we must perform the following substitutions for the fermionic fields
in the Lagrangian (Eq. (\ref{lagrangian})) of the model

\begin{figure}[t]
 \includegraphics[height=1.2cm]{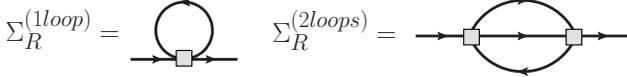}
 \caption{Feynman diagrams for the self-energy of the model up to two-loop order.
The one-loop contribution is the Hartree term and the two-loop contribution is the so-called sunset diagram.}
 \label{fig:Self_energy}
\end{figure}

\vspace{-0.4cm}

\begin{eqnarray}\label{Z}
\psi_{\sigma}(\mathbf{k})\rightarrow Z_{\Lambda}^{1/2}\psi_{R\sigma}(\mathbf{k}),\nonumber\\
\overline{\psi}_{\sigma}(\mathbf{k})\rightarrow Z^{1/2}_{\Lambda}\overline{\psi}_{R\sigma}(\mathbf{k}),
\end{eqnarray}

\noindent where $Z_{\Lambda}$ is the RG flowing quasiparticle weight which is
naturally related in the limit of $\Lambda\rightarrow 0$ to the conventional
many-body definition of the quasiparticle peak $Z=(1-\partial \text{Re}\Sigma(\omega,\mathbf{k}=\mathbf{k_F})/\partial\omega|_{\omega=0})^{-1}$,
with $\Sigma(\omega,\mathbf{k})$ being the self-energy and $\mathbf{k_F}$ is the Fermi vector.
The corresponding Feynman diagrams up to two-loop order are shown in Fig. \ref{fig:Self_energy}. The one-loop diagram
(i.e. the Hartree term) is generally independent of the external frequency. As a result, this contribution
does not renormalize the quasiparticle weight in the present case. In fact, it only generates a constant shift in
the chemical potential of the model that must be appropriately subtracted by a counterterm such that the
density of particles in the system remains always fixed\cite{Shankar}. By contrast, the two-loop contribution (i.e. the sunset diagrams)
is the first contribution to the self-energy that produces a non-analyticity as a function of the external
frequency $\omega$. For this reason, this term alone will be responsible for the renormalization
of the quasiparticle weight up to this order (for more details on this point, see, e.g., Refs. \cite{Freire2,Freire6} in the context of different fermionic 2D models).

The so-called anomalous dimension is conventionally defined by $\eta=\Lambda \,d\ln Z_{\Lambda}/d\Lambda$ (see, e.g., Ref. \cite{Peskin}). From this
expression, we obtain using a standard RG condition for the inverse of the renormalized single-particle Green's function $\Gamma^{(2)}_{R}=(G_{R})^{-1}$, i.e. 
$\text{Re}\,\Gamma^{(2)}_{R}(k_0=\Lambda,\mathbf{k}=\mathbf{k_F})=\Lambda$ that

\vspace{-0.3cm}

\begin{eqnarray}\label{eta}
\eta&=&\frac{1}{4}\bigg(g_{1}^{2}+g_{2}^{2}+\frac{g_{3}^{2}}{2}+g_{1c}^{2}+g_{2c}^{2}+g_{1x}^{2}+g_{2x}^{2}+g_{3x}^{2}\nonumber\\
&-&g_{1}g_{2}-g_{1c}g_{2c}-g_{1x}g_{2x}-g_{3p}g_{3x}+g_{3p}^{2}\bigg).
\end{eqnarray}

\begin{figure}[b]
 \includegraphics[height=4.3cm]{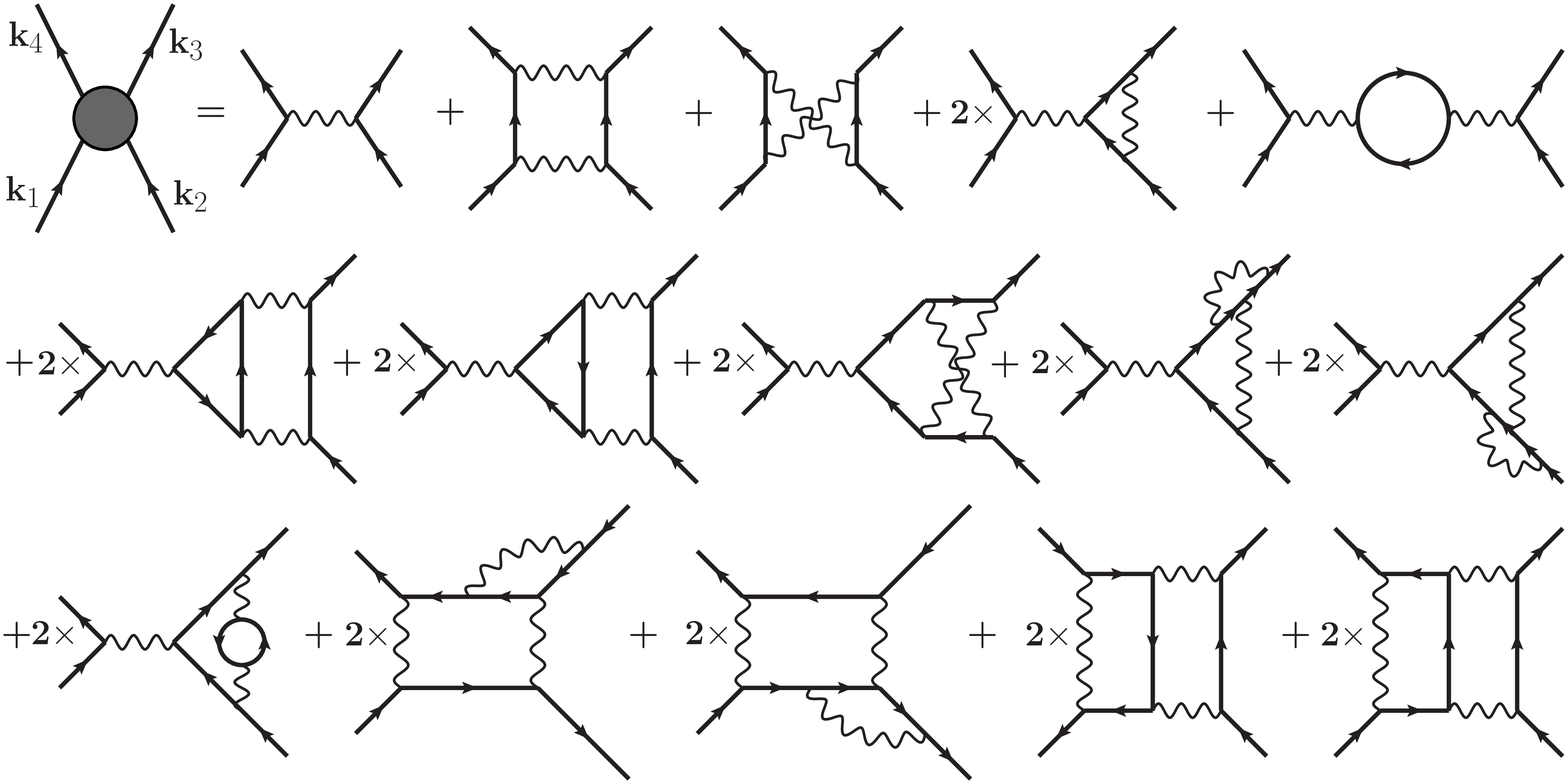}
 \includegraphics[height=1.3cm]{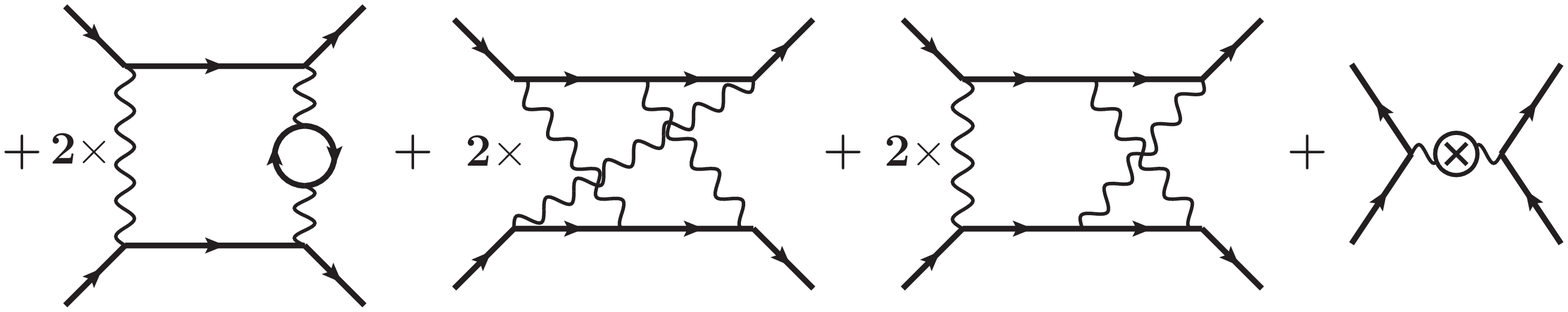}
 \caption{A representation of the most relevant Feynman diagrams for the vertex corrections up to two-loop order. The last diagram represents the counterterm.}
 \label{fig:Vertex_corrections}
\end{figure}

\noindent Here we would like to stress the fact that if we were to take the
1D limit of the above equation, we would reproduce exactly the
well-known result for the anomalous dimension for this case at two-loop order as was first calculated long
ago in Ref. \cite{Solyom}.

In addition, it can also be conventionally defined\cite{Peskin} that
the bare and renormalized coupling constants are related to
each other at two loops by

\vspace{-0.3cm}

\begin{eqnarray}\label{coupling}
g_{i,B}=N^{-1}(0)Z_{\Lambda}^{-2}\left[g_{iR}+\Delta g_{iR}^{1loop}+\Delta g_{iR}^{2loops}\right],
\end{eqnarray}

\noindent where $i=1,2,3,1c,2c,1x,2x,1s,1r,3x,3p,3t$ and the term $N(0)=k_c/(\pi^2 v_{F,\perp})$ is 
related to the density of states of the model at the ``hot spots'' points. The renormalized quantities --
labeled by the subscript $R$ -- generally depend on the RG scale $\Lambda$. In contrast, all the bare
quantities will be denoted by the label $B$. The terms $\Delta g_{iR}^{1loop}$ and $\Delta g_{iR}^{2loops}$
represent the counterterms necessary to regularize at one-loop and two-loop orders, respectively, the one-particle irreducible
four-point functions in each of the corresponding scattering channels.

We can now adjust these counterterms
such that all divergences are exactly canceled in our series
expansion for the couplings up to two loops, i.e. we choose the following standard RG condition for the one-particle irreducible
four-point vertices: $\Gamma_{i,R}^{(4)}(k_{10}=\Lambda/2,k_{20}=\Lambda/2,k_{30}=3\Lambda/2,k_{40}=-\Lambda/2)=-ig_{i,R}(\Lambda)$, where $k_{i0}$ are the energy
components. However, the price we pay for
this is the appearance of a new scale with all physical
quantities now depending on this scale $\Lambda$. By contrast, the
original model has no information about this quantity,
i.e. the bare parameters do not depend on $\Lambda$. This leads
us to the renormalization group conditions for the bare
couplings of the model, i.e. $\Lambda \,d g_{i,B}/d\Lambda=0$.
Therefore, using Eq. (\ref{coupling}), we finally obtain

\vspace{-0.3cm}

\begin{eqnarray}\label{g}
\Lambda\frac{dg_{i,R}}{d\Lambda}=2\eta\, g_{i,R}-\Lambda\frac{d}{d\Lambda}\left(\Delta g_{i,R}^{1loop}+\Delta g_{i,R}^{2loops}\right).
\end{eqnarray}

\noindent The initial conditions for this
system of differential equations are naturally given by the
microscopic interactions. A representation of the most relevant Feynman
diagrams corresponding to the vertex corrections up to
two-loop order are displayed in Fig. \ref{fig:Vertex_corrections}.

In order to investigate what are the enhanced correlations in the low-energy limit of the model, it is important to
calculate the corresponding susceptibilities by introducing
an infinitesimal external field in the appropriate channel
and evaluating its linear response. Therefore, we must
add to the Lagrangian that describes the present model the
following term

\begin{figure}[t]
 \includegraphics[height=4.9cm]{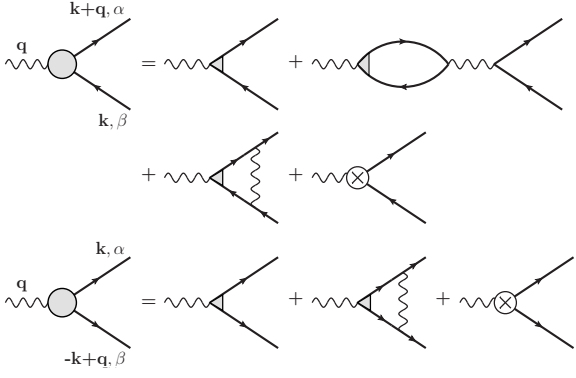}
 \caption{Feynman diagrams for both SC and DW response functions of the model. The diagrams with crosses represent the counterterms.}
 \label{fig:response_function}
\end{figure}

\vspace{-0.2cm}

\begin{align}\label{external}
&L_{ext}=\sum_{\mathbf{k},\alpha,\beta}\mathcal{T}_{B,SC}^{\alpha\beta}\,\overline{\psi}_{\alpha}(\mathbf{k})\overline{\psi}_{\beta}(-\mathbf{k}) \nonumber\\
&+\sum_{\mathbf{k},\alpha,\beta}\mathcal{T}_{B,DW}^{\alpha\beta}\,\overline{\psi}_{\alpha}(\mathbf{k}+\mathbf{Q})\psi_{\beta}(\mathbf{k})
+ H.c.,
\end{align}

\noindent where $\mathcal{T}_{B,SC}^{\alpha\beta}$ and $\mathcal{T}_{B,DW}^{\alpha\beta}$ are the bare response vertices
for superconducting (SC) and density-wave (DW) orders, respectively, and $\mathbf{Q}=(\pi,\pi)$. This
added term will generate new Feynman diagrams -- the three-legged vertices displayed in Fig. \ref{fig:response_function} -- which will also produce
new logarithmic singularities in the low-energy limit of our
field theory model (see also Refs. \cite{Freire2,Freire6} in the context of different fermionic 2D models). Therefore, we must regularize these
divergences by defining the renormalized response vertices and the corresponding counterterms as follows

\vspace{-0.2cm}

\begin{eqnarray}\label{response}
\mathcal{T}_{B,SC}^{\alpha\beta}&=& Z_{\Lambda}^{-1}\left[\mathcal{T}_{R,SC}^{\alpha\beta}(k_{\parallel},k_{\perp})+\Delta \mathcal{T}_{R,SC}^{\alpha\beta}(k_{\parallel},k_{\perp})\right],\nonumber\\
\mathcal{T}_{B,DW}^{\alpha\beta}&=& Z_{\Lambda}^{-1}\left[\mathcal{T}_{R,DW}^{\alpha\beta}(k_{\parallel},k_{\perp})+\Delta \mathcal{T}_{R,DW}^{\alpha\beta}(k_{\parallel},k_{\perp})\right].\nonumber\\
\end{eqnarray}

\noindent To calculate these counterterms, we must also establish a RG prescription 
$\Gamma^{(2,1)}_{R,i,\alpha\beta}(k_0=\Lambda,k_{\parallel},k_{\perp}=\pm k_{F,\perp})=-i\mathcal{T}_{R,i}^{\alpha\beta}(k_{\parallel},\pm k_{F,\perp};\Lambda)$
for $i=SC$ and $DW$.
As a result, by invoking the RG condition for bare quantities of the model $\Lambda (d\mathcal{T}^{\alpha\beta}_{B,SC(DW)}/d\Lambda) =0$,
we obtain the flow equations for the SC response vertices

\vspace{-0.2cm}

\begin{align}\label{SC}
&\Lambda\frac{d\mathcal{T}_{R,SC}^{(1)\alpha\beta}}{d\Lambda}=\frac{1}{2}\left[(g_{2c}+g_{1r})\mathcal{T}_{R,SC}^{(1)\alpha\beta}
-(g_{1c}+g_{1r})\mathcal{T}_{R,SC}^{(1)\beta\alpha}\right.\nonumber\\
&\left.+(g_{2x}+g_{1s})\mathcal{T}_{R,SC}^{(2)\alpha\beta}-(g_{1x}+g_{1s})\mathcal{T}_{R,SC}^{(2)\beta\alpha}\right]+\eta\,\mathcal{T}_{R,SC}^{(1)\alpha\beta},\nonumber\\
\end{align}

\noindent where we defined the new vertices $\mathcal{T}_{R,SC}^{(1)\alpha\beta}=\mathcal{T}_{R,SC}^{\alpha\beta}(k_{\parallel},k_{F,\perp})$
and $\mathcal{T}_{R,SC}^{(2)\alpha\beta}=\mathcal{T}_{R,SC}^{\alpha\beta}(k_{\parallel},-k_{F,\perp})$ depending on their location at the Fermi points. As for the DW response vertices we get
the following RG equations

\vspace{-0.2cm}

\begin{align}\label{DW}
&\Lambda\frac{d\mathcal{T}_{R,DW}^{(1)\alpha\beta}}{d\Lambda}=\frac{1}{2}\bigg[(g_1+g_3+2g_{3t})\sum_{\sigma=\alpha,\beta}\mathcal{T}_{R,DW}^{(1)\sigma\sigma}\nonumber\\
&+(g_{1x}+g_{3p}+2g_{3t})\sum_{\sigma=\alpha,\beta}\mathcal{T}_{R,DW}^{(2)\sigma\sigma}-(g_{3}+g_{3x})\mathcal{T}_{R,DW}^{(1)\alpha\beta}\nonumber\\
&-(g_{2}+g_{2x})\mathcal{T}_{R,DW}^{(2)\beta\alpha}\bigg]+\eta\,\mathcal{T}_{R,DW}^{(1)\alpha\beta},
\end{align}

\noindent where we also introduced the new vertices $\mathcal{T}_{R,DW}^{(1)\alpha\beta}=\mathcal{T}_{R,DW}^{\alpha\beta}(k_{\parallel},k_{F,\perp})$
and $\mathcal{T}_{R,DW}^{(2)\alpha\beta}=\mathcal{T}_{R,DW}^{\alpha\beta}(k_{\parallel},-k_{F,\perp})$ according to their association to the hot spots.
By symmetrizing and antisymmetrizing all the response vertices w.r.t. the spin indices, we obtain the following order parameters

\vspace{0.5cm}

$\left\{%
\begin{array}{ll}
    \mathcal{T}^{(j)}_{SSC}=\mathcal{T}_{R,SC}^{(j)\uparrow\downarrow}-\mathcal{T}_{R,SC}^{(j)\downarrow\uparrow},\\ \\
    \mathcal{T}^{(j)}_{TSC}=\left\{\begin{array}{ll}
                             \mathcal{T}_{R,SC}^{(j)\uparrow\uparrow}, \hspace{0.3cm} (S_z=1)\\
                             \mathcal{T}_{R,SC}^{(j)\uparrow\downarrow}+\mathcal{T}_{R,SC}^{(j)\downarrow\uparrow}, \hspace{0.3cm} (S_z=0)\\
                             \mathcal{T}_{R,SC}^{(j)\downarrow\downarrow}, \hspace{0.3cm} (S_z=-1)\\
                             \end{array}\right.\\ \\
    \mathcal{T}^{(j)}_{CDW}=\mathcal{T}_{R,DW}^{(j)\uparrow\uparrow}+\mathcal{T}_{R,DW}^{(j)\downarrow\downarrow},\\ \\
    \mathcal{T}^{(j)}_{SDW}=\mathcal{T}_{R,DW}^{(j)\uparrow\uparrow}-\mathcal{T}_{R,DW}^{(j)\downarrow\downarrow},,
\end{array}%
\right.$

\vspace{0.5cm}

\noindent where $j=1,2$ and the subscripts $SSC$ and $TSC$ correspond to singlet and triplet superconductivity,
whereas $SDW$ and $CDW$ stand for charge and spin-density waves, respectively. Hence, using the above relations
one can derive in a straightforward way the RG flow equations for each response vertex associated with a potential instability of the
normal state toward a given ordered (i.e. symmetry-broken) phase. In order to determine the symmetry of
the order parameter we must further symmetrize the response vertices w.r.t. the index $j$. Thus

\vspace{0.5cm}

$\left\{%
\begin{array}{ll}
    \mathcal{T}^{(s-wave)}_{SSC}=\mathcal{T}_{SSC}^{(1)}+\mathcal{T}_{SSC}^{(2)}, \vspace{0.1cm} \\
    \mathcal{T}^{(d-wave)}_{SSC}=\mathcal{T}_{SSC}^{(1)}-\mathcal{T}_{SSC}^{(2)},\vspace{0.1cm} \\
    \mathcal{T}^{(p-wave)}_{TSC}=\mathcal{T}_{TSC}^{(1)}+\mathcal{T}_{TSC}^{(2)},\vspace{0.1cm} \\
    \mathcal{T}_{CDW}=\mathcal{T}_{CDW}^{(1)}+\mathcal{T}_{CDW}^{(2)},\vspace{0.1cm} \\
    \mathcal{T}_{SDW}=\mathcal{T}_{SDW}^{(1)}+\mathcal{T}_{SDW}^{(2)},\vspace{0.1cm} \\
    \mathcal{T}_{CF}=\mathcal{T}_{CDW}^{(1)}-\mathcal{T}_{CDW}^{(2)},\vspace{0.1cm} \\
    \mathcal{T}_{SF}=\mathcal{T}_{SDW}^{(1)}-\mathcal{T}_{SDW}^{(2)}.\\
\end{array}%
\right.$

\vspace{0.5cm}

\noindent Here the labels $s-wave$, $d-wave$ and $p-wave$ refer to the symmetries of the corresponding SC order
parameters, and $CF$ and $SF$ stand for charge and spin flux phases, respectively.

\begin{figure}[t]
 \includegraphics[height=6.3cm]{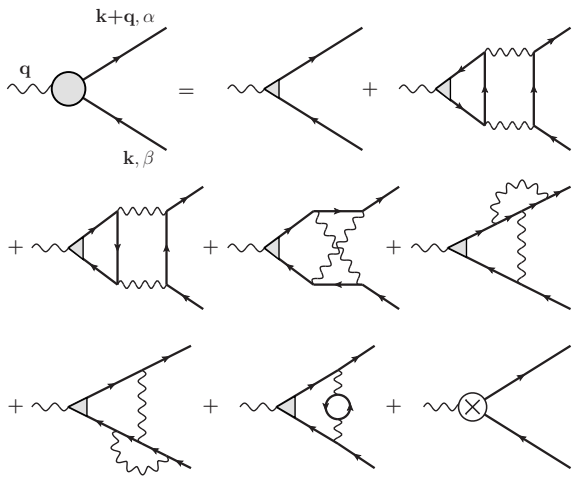}
 \caption{Feynman diagrams for uniform ($\mathbf{q}\rightarrow 0$) response functions of the model up to two loops. Since
the corresponding one-loop diagrams for this case are non-divergent near the hot spots, 
they turn out not to be relevant in this calculation. The last diagram stands for the counterterm.}
 \label{fig:uniform_response}
\end{figure}

We now move on to discuss the uniform ($\mathbf{q}\rightarrow 0$) linear response functions of the model. At one-loop level, we find that there
are no Feynman diagrams that produce logarithmic singularities in the low-energy limit. In this way,
it becomes necessary to go up to two loops in order to obtain the appropriate RG equations for this
quantity (see Fig. \ref{fig:uniform_response}). To do this, we must add new bare auxiliary fields $\mathcal{T}^{\alpha\alpha}_{B}(\mathbf{q\approx 0})$ that couple, respectively,
to the charge and spin densities in the Lagrangian of the model as follows

\vspace{-0.2cm}

\begin{align}\label{external}
&L'_{ext}=\sum_{\mathbf{k}}\big[\mathcal{T}^{\uparrow\uparrow}_{B}(\mathbf{q})\overline{\psi}_{\uparrow}(\mathbf{k})\psi_{\uparrow}(\mathbf{k})+
\mathcal{T}^{\downarrow\downarrow}_{B}(\mathbf{q})\overline{\psi}_{\downarrow}(\mathbf{k})\psi_{\downarrow}(\mathbf{k})\big].
\end{align}

\noindent Thus, we must regularize these divergences by defining the renormalized uniform response functions
and the corresponding counterterms as follows\cite{Freire3,Freire4}

\vspace{-0.4cm}

\begin{align}
\mathcal{T}_{B}^{\alpha\alpha}(\mathbf{q})&=Z^{-1}_{\Lambda}\big[\mathcal{T}_{R}^{\alpha\alpha}(\mathbf{q})+\Delta \mathcal{T}_{R}^{\alpha\alpha}(\mathbf{q})\big],
\end{align}

\noindent where the $Z_{\Lambda}$ factor originates, as discussed previously, from the redefinition of
the fermionic fields at two-loop RG level displayed in Eq. (\ref{Z}) and is naturally related
to the self-energy feedback into the equations. In order to determine these counterterms, we 
must change the RG condition of the three-legged vertices to 
$\Gamma^{(2,1)}_{R,\alpha\alpha}(k_0=0, v_{F,\perp}(k_{\parallel}) (|k_{\perp}|-k_{F,\perp})=\Lambda,\mathbf{q}\approx 0)=-i\mathcal{T}^{\alpha\alpha}_{R}(\mathbf{q};\Lambda)$ since we
will now approach the low-energy limit of the model via the momentum scale.
Following this prescription, by establishing the RG condition for the bare quantities of the model $\Lambda (d\mathcal{T}_{B}^{\alpha\alpha}/d\Lambda) =0$,
we obtain the RG equations for the renormalized uniform response functions

\vspace{-0.3cm}

\begin{align}
\Lambda\frac{d\mathcal{T}_{R}^{\alpha\alpha}(\mathbf{q})}{d\Lambda}&=\eta\mathcal{T}_{R}^{\alpha\alpha}(\mathbf{q})-\Lambda\frac{d\Delta\mathcal{T}_{R}^{\alpha\alpha}(\mathbf{q})}{d\Lambda}.
\end{align}

\noindent By symmetrizing and antisymmetrizing these quantities w.r.t. spin indices, i.e.

\vspace{+0.3cm}

$\left\{%
\begin{array}{ll}
    \mathcal{T}_{C}=\mathcal{T}_{R}^{\uparrow\uparrow}+\mathcal{T}_{R}^{\downarrow\downarrow}, \vspace{0.1cm} \\
    \mathcal{T}_{S}=\mathcal{T}_{R}^{\uparrow\uparrow}-\mathcal{T}_{R}^{\downarrow\downarrow}, 
\end{array}%
\right.$

\vspace{+0.3cm}

\noindent we readily obtain the corresponding flow equations for both renormalized charge ($\mathcal{T}_{C}$) and spin ($\mathcal{T}_{S}$) uniform response functions.

Lastly, once we computed the response vertices associated with all the order parameters,
we can proceed to calculate their corresponding renormalized susceptibilities. They
are given by

\vspace{-0.5cm}

\begin{equation}\label{susc}
\chi_{m}(\Lambda)=N(0)\int_{0}^{l} d\xi\mathcal{T}_{m}(\xi)\mathcal{T}^{*}_{m}(\xi),
\end{equation}

\noindent where $m=SSC(s-wave),SSC(d-wave),TSC(p-wave),CDW,SDW,CF$ and $SF$, and the RG step $l$ is defined by $l=\ln(\Lambda_0/\Lambda)$.
As for the uniform susceptibilities, one obtains that $\chi_{n}(\Lambda)=N(0)|\mathcal{T}_{n}(\Lambda)|^{2}$, where $n=C$ and $S$.

\section{Two-loop RG Flow Equations and Numerical Solution}

In this section, we discuss the RG flow equations up to two-loop order associated with all the physical quantities calculated previously in this paper and 
show their numerical solution as one approaches the low-energy limit of the model. Concerning the renormalized couplings, 
we choose to write down for completeness all the RG equations up to two-loop order explicitly in Appendix A. 
It is important to stress here that, up to one-loop order, our equations agree with 
those derived by Furukawa and Rice (Ref. \cite{Rice}) who first discussed a very similar model within a one-loop RG approximation in the literature. 
However, as we have explained previously in this paper, to go beyond the work in Ref. \cite{Rice}
we need to investigate also the effect of higher-order quantum fluctuations in order to describe their impact on the low-energy dynamics of the system.

\begin{figure}[b]
 \includegraphics[height=2.2in]{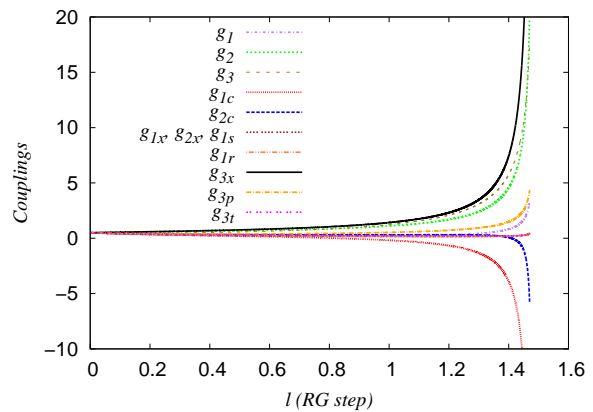}
 \caption{(Color online) One-loop RG flow for all the coupling constants of the model for the initial choice of $g_i^{(0)}=0.5$.}
 \label{fig:RG_couplings}
\end{figure}

To solve all the coupled differential RG equations numerically, we apply standard
fourth-order Runge-Kutta method. As an initial condition for these equations, we take
Hubbard-like repulsive parameters, i.e. $g_i^{(0)}=[k_c/(\pi^2 v_{F,\perp})]\,U$ for all couplings. For simplicity, we will always choose 
$g_i^{(0)}=0.5$ in the present work. 
Using the experimental data from Ref. \cite{Campuzano}, since the nodal liquid is stabilized for $4\%$ of hole doping, this implies that 
$v_{F,\perp}\approx 1.70t$ at
the ``hot spots'' and the Hubbard local interaction parameter $U\approx 3.6t$. We mention here, however, that our results do not depend
crucially on this initial choice of the coupling constants and we emphasize that they are robust within an appreciable range of these physical parameters.

In order to establish a direct comparison between the one-loop RG results and the two-loop RG data for this 2D model, we first
present a plot of the flow of 
the couplings up to one-loop order only. As a result, the corresponding one-loop RG flow
as a function of the step $l$ is displayed in Fig. \ref{fig:RG_couplings}. 
In agreement with Ref.\cite{Rice}, we confirm that despite the fact that the couplings are initially taken to be weak,
the one-loop RG flow is always to a strong coupling regime in the low-energy limit for Hubbard-like initial interactions. Additionally, we
observe that many renormalized couplings tend to
diverge as the RG scale is lowered at the same critical scale $l_c=\ln(\Lambda_0/\Lambda_c)$, after which point the RG approach up
to one loop clearly breaks down. This fact seems to limit
the validity of the one-loop RG calculation in order to study
the infrared regime of this model and invites one to go beyond
that approximation in order to access the low-energy limit of the system.

\begin{figure}[t]
 \includegraphics[height=2.2in]{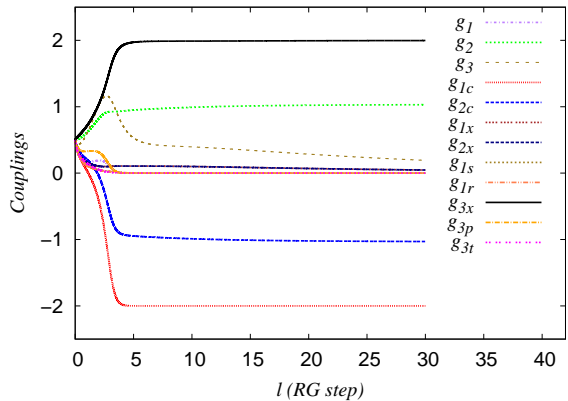}
 \caption{(Color online) Two-loop RG flow for all the coupling constants of the model for the initial choice of $g_i^{(0)}=0.5$.}
 \label{fig:RG_couplings_two}
\end{figure}

Following this strategy, we now move on to our two-loop RG results. The RG flow up to two-loop order for the couplings as a function of the step $l$ is  
depicted in Fig. \ref{fig:RG_couplings_two}. In this plot, we observe that instead of displaying a divergent behavior 
as shown in the one-loop RG flow, all renormalized couplings at two-loop RG level
now clearly approach asymptotically infrared stable fixed points in the low-energy limit. 
It is true, however, that some couplings become saturated at reasonably strong coupling
fixed points. This is a well-known problem in
RG theory and happens as well in other applications of
this method to quantum field theories in which fluctuation
effects are known to be strong (the most notorious
example being the Wilson-Fisher fixed point
in $\phi^{4}$--theory in three dimensions\cite{Wilson,Brezin}).
Notwithstanding this fact, we point out that, quite surprisingly, the two-loop RG approach turns out to yield a better controlled
theory for the present model than the one-loop RG scheme   
and, for this reason, we can hope that it could describe at least qualitatively
the correct trend of the low-energy dynamics of the system.
Two increasingly relevant renormalized couplings in the RG flow turn out to be the Umklapp scattering $g_{3x}$ and the forward scattering $g_2$. As we will
see shortly, they both tend to favor SDW antiferromagnetic ordering tendencies in the system. 
By contrast, upon inclusion of quantum fluctuation effects, the Cooper pair interaction processes $g_{1c}$ and $g_{2c}$ -- that are initially taken to be 
repulsive -- flow naturally to attractive infrared stable fixed points in the low-energy limit. These latter
couplings in turn enhance SC pairing correlations which will of course tend to manifest itself as a possible competing order in the system.
As can also be inferred from Fig. \ref{fig:RG_couplings_two}, the remaining couplings flow asymptotically to zero in the infrared regime
and, as a result, they become (dangerously) irrelevant in the low-energy effective description of this 2D model.

\begin{figure}[b]
 \includegraphics[height=2.2in]{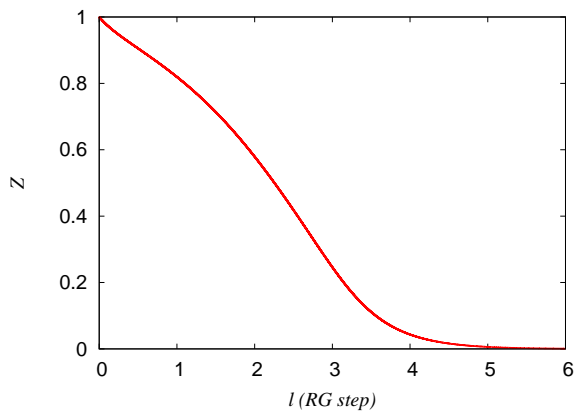}
 \caption{(Color online) Two-loop RG flow for the quasiparticle weight $Z_{\Lambda}$ of the model for the initial choice of $g_i^{(0)}=0.5$.}
 \label{fig:RG_quasiparticle}
\end{figure}

Next, we focus on the RG flow equation up to two-loop order for the quasiparticle weight $Z_{\Lambda}$ of the model (see Eq. (\ref{eta})). By solving
this equation numerically with initial condition $Z_{\Lambda}(l=0)=1$, we obtain the results depicted in the plot of Fig. \ref{fig:RG_quasiparticle} 
as a function of the RG step $l$. We observe that this quantity becomes universally
suppressed as a power-law $Z_{\Lambda}\sim(\Lambda/\Lambda_0)^{\eta}$ with $\eta\approx 2.01$ in the low-energy limit, 
indicating a complete absence of coherent quasiparticles and a
clear breakdown of Fermi liquid behavior near the ``hot spots'' 
for this regime. This conclusion is also in qualitative agreement with other works available in the literature \cite{Abanov,Abanov2,Sachdev}. 
Thus, in order to investigate the true nature of this low-energy state, 
we must further characterize the model by examining what are the dominant fluctuations that
drive the system to a nullified $Z_\Lambda$ in the low-energy limit.

For this reason, we now turn our attention to the behavior of the various order-parameter
susceptibilities as a function of the RG scale $\Lambda$ of the model. Using Eqs. (\ref{SC}) and (\ref{DW}), we write
down explicitly the corresponding RG flow equations for these quantities up to two-loop order as follows

\vspace{-0.2cm}

\begin{eqnarray}
\dot{\mathcal{T}}_{SSC}^{s-wave}&=&\frac{1}{2}(g_{2c}+g_{1c}+g_{1x}+g_{2x}+2g_{1s}+2g_{1r})\nonumber\\
&\times&\mathcal{T}_{SSC}^{s-wave}+\eta\,\mathcal{T}_{SSC}^{s-wave},\\
\dot{\mathcal{T}}_{SSC}^{d-wave}&=&\frac{1}{2}(g_{2c}+g_{1c}-g_{1x}-g_{2x}-2g_{1s}+2g_{1r})\nonumber\\
&\times&\mathcal{T}_{SSC}^{d-wave}+\eta\,\mathcal{T}_{SSC}^{d-wave},\\
\dot{\mathcal{T}}_{TSC}^{p-wave}&=&\frac{1}{2}(g_{2c}-g_{1c}+g_{2x}-g_{1x})\mathcal{T}_{TSC}^{p-wave}\nonumber\\
&+&\eta\,\mathcal{T}_{TSC}^{p-wave},\\
\dot{\mathcal{T}}_{CDW}&=&\frac{1}{2}(2g_{1}+g_{3}+2g_{1x}+2g_{3p}+8g_{3t}\nonumber\\
&-&g_{3x}-g_{2}-g_{2x})\mathcal{T}_{CDW}+\eta\,\mathcal{T}_{CDW},\\
\dot{\mathcal{T}}_{SDW}&=&-\frac{1}{2}(g_{2}+g_{2x}+g_{3}+g_{3x})\mathcal{T}_{SDW}\nonumber\\
&+&\eta\,\mathcal{T}_{SDW},\\
\dot{\mathcal{T}}_{CF}&=&\frac{1}{2}(2g_{1}+g_{2}-2g_{1x}+g_{2x}+g_{3}\nonumber\\
&-&2g_{3p}-g_{3x})\mathcal{T}_{CF}+\eta\,\mathcal{T}_{CF},\\
\dot{\mathcal{T}}_{SF}&=&-\frac{1}{2}(g_{3}+g_{3x}-g_{2}-g_{2x})\mathcal{T}_{SF}\nonumber\\
&+&\eta\,\mathcal{T}_{SF},
\end{eqnarray}

\begin{figure}[t]
 \includegraphics[height=2.2in]{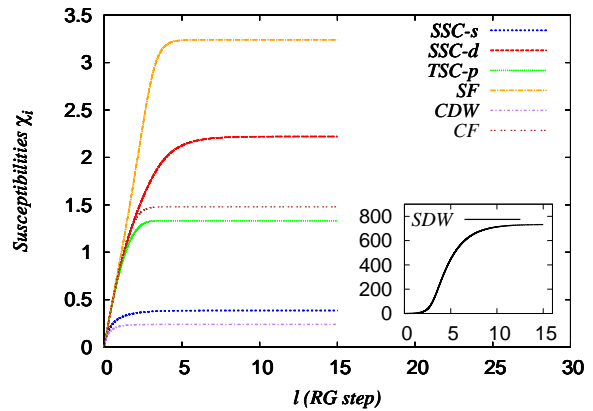}
 \caption{(Color online) Two-loop RG flow for various order-parameter susceptibilities of the model (in units of $N(0)$) for the initial choice of $g_i^{(0)}=0.5$.}
 \label{fig:RG_susceptibilities}
\end{figure}

\noindent where we are using the shorthand notation $\dot{\mathcal{T}_{i}}=\Lambda(d\mathcal{T}_{i})/(d\Lambda)$.
We solve these differential equations using the same numerical procedure explained above and, by using Eq. (\ref{susc}), we are 
able to follow the RG flow of the corresponding susceptibilities
as a function of the RG step $l$. The results are plotted in Fig. \ref{fig:RG_susceptibilities}. In this figure, we obtain that 
even though many susceptibilities are renormalized due to interactions, they remain non-divergent and
become saturated at plateaus in the low-energy limit at two loops. 
In other words, the self-energy feedback and the higher-order vertex corrections at two-loop RG level are sufficient to
suppress the divergence of these physical quantities which would show up in one-loop RG calculations.
Our present two-loop RG result implies that there
should be no spontaneous symmetry breaking associated with these renormalized order-parameter susceptibilities investigated here in this work and therefore
the system is not expected to exhibit long-range order in any of these channels at low energies. 
Indeed, despite the fact that the leading short-range correlations -- by at least two orders of magnitude --   
for this regime are of SDW-type (i.e. antiferromagnetic spin correlations), spin flux phase fluctuations also appear as a subleading ordering tendency 
and $d-wave$ singlet SC clearly comes in third place.
This suggests that a possible coexistence of such short-ranged ordering tendencies 
could be the hallmark of the true nature of the corresponding low-energy state. 

Other physical quantities which are naturally of high importance to potentially identify the nature of the ground state of the model are the 
uniform ($\mathbf{q}\rightarrow 0$) susceptibilities in both spin and charge sector. After a lengthy but straightforward calculation, we obtain
that the resulting RG flow equations for
the uniform response functions up to two-loop order are the following

\vspace{-0.3cm}

\begin{align}\label{Unif_resp}
\dot{\mathcal{T}}_{S}&=\frac{1}{2}\big[g_{1}^{2}+g_{1x}^{2}+g_{1c}^{2}+(g_{3p}-g_{3x})^2\big]\mathcal{T}_{S},\nonumber\\
\dot{\mathcal{T}}_{C}&=\frac{1}{2}\big[g_{3}^{2}+g_{3p}^{2}+g_{3x}^{2}+(g_{3p}-g_{3x})^2\big]\mathcal{T}_{C},
\end{align}

\noindent where $\dot{\mathcal{T}_{i}}=\Lambda(d\mathcal{T}_{i})/(d\Lambda)$.  We emphasize here the fact that 
if we take, as an important check, the 1D limit of the above two-loop RG equations for
the uniform spin and charge response functions, we are able to reproduce established results available in the literature for these systems 
in various regimes (see, e.g., Refs. \cite{Solyom,Fukuyama,Voit}).

\begin{figure}[t]
 \includegraphics[height=2.1in]{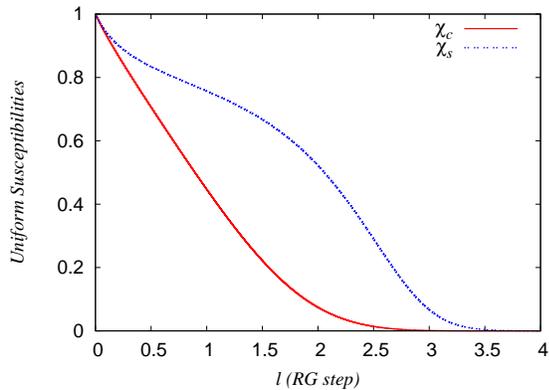}
 \caption{(Color online) Two-loop RG flow for both charge (C) and spin (S) uniform susceptibilities of the model (in units of $N(0)$) for the initial choice of $g_i^{(0)}=0.5$.}
 \label{fig:uniform_susceptibilities}
\end{figure}

In the present 2D model, we once more follow the 
numerical procedure described before in order to solve the equations given by Eqs. (\ref{Unif_resp}) and Eq. (\ref{susc}). 
Our result for the uniform charge susceptibility 
is shown in Fig. \ref{fig:uniform_susceptibilities}.
We observe from this plot that, for repulsive Hubbard-like initial conditions, the uniform charge 
susceptibility (or simply charge compressibility) becomes strongly suppressed and renormalizes to zero in the low-energy limit.
This implies that there should be a gap opening in the charge excitation spectrum in the system. This
charge gap is produced by the Umklapp coupling parameter $g_{3x}$, which, as we have seen before, becomes increasingly relevant in the low-energy
effective theory. At this point, we draw attention to the fact that our result bears some resemblance to
the well-studied case of the two-chain repulsive Hubbard ladder exactly at half-filling\cite{Balents3,Balents4}, where it is by now well-established that 
the mechanism for charge gap generation in that system is driven by Umklapp interaction
in the scaling limit. Therefore, we may conclude from this that the nature of the low-energy state of the present 2D model is clearly insulating, 
which also agrees with our previous result that
the quasiparticle weight $Z_{\Lambda}$ is nullified in the low-energy limit.  

In Fig. \ref{fig:uniform_susceptibilities}, we also display the RG flow for the uniform spin susceptibility of the system for 
repulsive Hubbard-like initial conditions. As a result, 
we observe from this plot that even though this quantity is initially less suppressed in the RG flow than the charge compressibility, 
it also scales down eventually to zero in 
the low-energy limit. This indicates that there should be also a gap in the spin excitation spectrum of the model. The opening of a spin gap is 
produced by an interplay  
between the Umklapp coupling parameter $g_{3x}$ (which also promotes SDW antiferromagnetic fluctuations and leads to charge gap formation) 
and the Cooper pair interaction process $g_{1c}$,
which in turn renormalizes to an attractive infrared stable fixed point at low energies and enhances $d-wave$ pairing correlations. 
This result strongly hints at pronounced but short-ranged antiferromagnetic spin correlations coexisting with short-ranged $d-wave$
singlet superconductivity as the underlying physics 
of this corresponding low-energy state.

Hence, it seems reasonable to suppose that our two-loop RG data  
could suggest that the low-energy state of the present 2D model 
may indeed share some similarities to the $d$-Mott phase of the half-filled two-chain repulsive 
Hubbard ladder where the spin gap 
is well-understood to be related to a precursor effect of ``pre-formed'' Cooper pairs with approximate $d-wave$ symmetry in an 
otherwise insulating system \cite{Balents3,Balents4}. This interpretation
is in line with Ref. \cite{Rice}, where the authors also concluded, based on a one-loop RG calculation,
that the low-energy state of this 2D model should be given by an insulating spin-gapped state. 
However, we point out here that, within a RG computation up to one-loop order only, one cannot be sure about the spin properties
of this state since the uniform spin susceptibility and the SDW spin susceptibility actually behave in a
contradictory manner \cite{Rice}. Our present two-loop RG results resolve this contradiction in a most natural way.
Since the SDW susceptibility at two-loop RG level becomes saturated at a plateau at low energies, it cannot be associated to
any long-range spin order. This is also consistent with the fact that the low-energy state naturally displays a 
spin gap. Therefore, we conclude that such an insulating spin-gapped state could be 
produced by quantum disordering effects induced by correlations included in the present 2D model, and the universal properties of this state
are captured qualitatively by our two-loop RG scheme within a weak-to-moderate coupling regime.

\section{Conclusions}

We have performed a RG calculation of a 2D model consisting of eight points located near the ``hot spots'' on a Fermi surface
which are directly connected by SDW ordering wave vector. By following the field-theoretical RG strategy, we have derived the corresponding flow equations 
for the couplings, the quasiparticle weight, the order-parameter susceptibilities and
the uniform spin and charge susceptibilities of the model up to two-loop order.
As a result, we have found for physically relevant choices of parameters that while SDW antiferromagnetism, spin flux phase, and $d$-wave 
superconductivity manifest themselves 
as coexisting short-range ordering tendencies, the quasiparticle weight $Z_{\Lambda}$ 
renormalizes to zero in the low-energy limit, indicating the breakdown of the Fermi liquid behavior for this regime. 
Moreover, both uniform spin and charge
susceptibilities scale down to zero in the same limit, which point to the fact that gap 
openings should take place in both spin and charge excitation spectra of the model. 
By comparing those results with other well-studied systems such as the half-filled two-chain repulsive Hubbard ladder, we
have shown that our two-loop RG data are consistent with an interpretation
that the low-energy state of the present 2D model should be given by an insulating spin-gapped state
displaying no long-range antiferromagnetic order at zero temperature. 

On the other hand, we would like to emphasize here that, despite the fact that the present 2D model might have a few analogies with 
the well-known physics of the pseudogap phase of the cuprate superconductors, 
this model must be of course extended in order to describe the complete Fermi surface (and, consequently, 
its process of either becoming partly truncated or fully reconstructed in momentum space) 
displayed by these compounds. For this reason, we believe that 
the present two-loop RG result could potentially give some insights to
this problem from a weak-to-moderate coupling perspective. 
Lastly, we point out that it would be naturally very interesting to perform such a 
complete two-loop RG analysis  
to discuss the fully 2D $t-t'$ Hubbard model on a square lattice with a choice of parameters that 
precisely match the Fermi surface observed experimentally in these materials.
This will introduce some expected complications such as the substantial increase of the renormalized 
couplings as they become in this case functions of three independent momenta of the 
interacting particles at low energies, but this can be handled numerically using standard techniques. Therefore, we plan
to perform such a RG investigation of the 2D $t-t'$ Hubbard model in another publication\cite{Freire5}.

\begin{acknowledgments}
We acknowledge financial support from CNPq and FAPEG for this project.
\end{acknowledgments}

\appendix

\section{}

In this appendix, we show explicitly the RG flow equations up to two loops for the model. They read

\begin{widetext}

\begin{eqnarray}
\Lambda\frac{dg_{1}}{d\Lambda}&=&g^{2}_{1}+g^{2}_{1x}+4g^{2}_{3t}+g^{2}_{3p}-g_{1x}g_{2x}-g_{3p}g_{3x}+\frac{1}{2}(g_{1x}g_{2x}-g^{2}_{2x}-g_{3p}g_{3x})g_{1c}\nonumber\\
&+&\frac{1}{2}(g^{2}_{1c}+g^{2}_{1}+g^{2}_{1x}+g^{2}_{2x}-g_{1x}g_{2x}-g_{3p}g_{3x}+g^{2}_{3p}+g^{2}_{3x})g_{1},\\
\Lambda\frac{dg_{2}}{d\Lambda}&=&\frac{1}{2}\left(g^{2}_{1}-g^{2}_{2x}-g^{2}_{3}-g^{2}_{3x}\right)+\frac{1}{4}\left(g^{3}_{1}+g_{1c}g^{2}_{1x}+g_{1}g^{2}_{1c}\right)+\frac{1}{4}\left(2g_{2}-g_{1}\right)g^{2}_{3}\nonumber\\
&+&\frac{1}{4}\left[\left(2g_{2c}-g_{1c}\right)\left(g^{2}_{3p}+g^{2}_{3x}\right)-2g_{2c}g_{3p}g_{3x}+2g_{2}\left(g^{2}_{3p}+g^{2}_{3x}-g_{3p}g_{3x}\right)\right]\nonumber\\
&+&\frac{1}{2}(g_{2}-g_{2c})(g_{1x}^{2}+g_{2x}^{2}-g_{1x}g_{2x}),\\
\Lambda\frac{dg_{3}}{d\Lambda}&=&(g_{1}-2g_{2})g_{3}+g^{2}_{3t}-g_{1x}(g_{3x}-2g_{3p})-g_{2x}(g_{3p}-g_{3x})+\frac{1}{4}[(g_{1}-2g_{2})^{2}\nonumber\\
&+&(g_{1c}-2g_{2c})^{2}+2g^{2}_{1x}+2g^{2}_{2x}-2g_{1x}g_{2x}-2g_{3p}g_{3x}+2g^{2}_{3p}+2g^{2}_{3x}+g^{2}_{3}]g_{3},\\
\Lambda\frac{dg_{1c}}{d\Lambda}&=&g^{2}_{1c}+g_{1x}g_{2x}+g^{2}_{1s}+g^{2}_{1r}+g^{2}_{3x}-g_{3p}g_{3x}+\frac{1}{2}\left(g_{1x}g_{2x}-g^{2}_{2x}-g_{3p}g_{3x}\right)g_{1}\nonumber\\
&+&\frac{1}{2}\left(g^{2}_{1c}+g^{2}_{1}+g^{2}_{1x}+g^{2}_{2x}-g_{1x}g_{2x} +g^{2}_{3p}+g^{2}_{3x}-g_{3p}g_{3x}\right)g_{1c},\\
\Lambda\frac{dg_{2c}}{d\Lambda}&=&\frac{1}{2}\left(g^{2}_{1c}+g^{2}_{1x}+g^{2}_{2x}+g^{2}_{1s}+g^{2}_{1l}+2g^{2}_{1r}-g^{2}_{3p}\right)+\frac{1}{4}\left(g_{1}g^{2}_{1x}+g_{1c}g^{2}_{1}+g^{3}_{1c}\right)\nonumber\\
&+&\frac{1}{2}(g_{2c}-g_{2})(g_{1x}^{2}+g^{2}_{2x}-g_{1x}g_{2x})+\frac{1}{4}\left(2g_{2c}-g_{1c}\right)g^{2}_{3}\nonumber\\
&+&\frac{1}{4}\left[\left(2g_{2}-g_{1}\right)\left(g^{2}_{3p}+g^{2}_{3x}\right)-2g_{2}g_{3p}g_{3x}+2g_{2c}\left(g^{2}_{3p}+g^{2}_{3x}-g_{3p}g_{3x}\right)\right],\\
\Lambda\frac{dg_{1x}}{d\Lambda}&=&g_{1c}g_{2x}+g_{2c}g_{1x}+2g_{1s}g_{1r}+\frac{1}{2}\biggl(g_{1}g_{2c}+g_{1c}g_{2}-2g_{2c}g_{2}-\frac{g^{2}_{3p}}{2}-\frac{g^{2}_{3x}}{2}\biggr)g_{1x}\nonumber\\
&+&2\eta~g_{1x},\\
\Lambda\frac{dg_{2x}}{d\Lambda}&=&g_{1c}g_{1x}+g_{2c}g_{2x}+2g_{1s}g_{1r}+\frac{1}{2}\biggl(g_{1c}g_{1}g_{1x}-2g_{1c}g_{1}g_{2x}+g_{1c}g_{2}g_{2x}\nonumber\\
&+&g_{1}g_{2c}g_{2x}-2g_{2c}g_{2}g_{2x}-g_{1x}g_{3p}g_{3x}+g_{2x}g_{3p}g_{3x}-\frac{1}{2}g_{2x}(g_{3x}^{2}+g_{3p}^{2})\biggr)+2\eta g_{2x},\\
\Lambda\frac{dg_{1s}}{d\Lambda}&=&(g_{1c}+g_{2c})g_{1s}+(g_{1x}+g_{2x})g_{1r}+2\eta g_{1s},\\
\Lambda\frac{dg_{1r}}{d\Lambda}&=&(g_{1c}+g_{2c})g_{1r}+\frac{1}{2}(g_{1x}+g_{2x})g_{1s}+2\eta g_{1r},\\
\Lambda\frac{dg_{3p}}{d\Lambda}&=&(2g_{1}-g_{2c})g_{3p}+g_{1x}g_{3}+4g^{2}_{3t}-g_{2}g_{3p}-g_{2x}g_{3}-g_{1}g_{3x}+\frac{1}{2}\biggl(2g_{2c}g_{2}g_{3p}\nonumber\\
&+&g^{2}_{2x}g_{3x}-g_{1}g_{2c}g_{3p}-g_{1c}g_{2}g_{3p}-g_{1x}g_{2x}g_{3x}-g_{1c}g_{1}g_{3x}-\frac{g^{2}_{1x}g_{3p}}{2}\biggr)+2\eta g_{3p},\\
\Lambda\frac{dg_{3x}}{d\Lambda}&=&(2g_{1c}-g_{2c})g_{3x}-g_{1c}g_{3p}-g_{2x}g_{3}-g_{2}g_{3x}+\frac{1}{2}\biggl(2g_{2c}g_{2}g_{3x}+g^{2}_{2x}g_{3p}-g_{1}g_{2c}g_{3x}\nonumber\\
&-&g_{1c}g_{2}g_{3x}-g_{1x}g_{2x}g_{3p}-g_{1c}g_{1}g_{3p}-\frac{g^{2}_{1x}g_{3x}}{2}\biggr)+2\eta g_{3x},\\
\Lambda\frac{dg_{3t}}{d\Lambda}&=&(2g_{1}-g_{2}+g_{3}+2g_{1x}-g_{2x}+2g_{3p}-g_{3x})g_{3t}+2\eta g_{3t},
\end{eqnarray}

\end{widetext}

\noindent where $\Lambda$ is the corresponding RG flowing scale and $\eta$ is naturally the anomalous dimension defined in Eq. (\ref{eta}). As we have discussed 
previously in this paper, the anomalous dimension is related to the self-energy feedback into the two-loop RG flow equations.

\end{document}